\documentclass{osa-article}

\journal{oe}

\articletype{Research Article}

\usepackage{lineno}
\usepackage{algorithm}
\usepackage{algorithmic}
\usepackage{amsmath}
\usepackage{makecell}
\usepackage[table]{xcolor}
\usepackage{nicematrix}

\begin{document}

\title{Analytical theory of frequency-modulated combs: generalized mean-field theory, complex cavities, and harmonic states}

\author{Levi Humbard and David Burghoff\authormark{1}}

\address{\authormark{1}Department of Electrical Engineering, University of Notre Dame, 257 Fitzpatrick Hall, Notre Dame, IN 46556, USA}

\email{\authormark{}lhumbard@nd.edu} 



\begin{abstract}
Frequency-modulated (FM) combs with a linearly-chirped frequency and nearly constant intensity occur naturally in certain laser systems; they can be most succinctly described by a nonlinear Schrödinger equation with a phase potential. In this work, we perform a comprehensive analytical study of FM combs in order to calculate their salient properties. We develop a general procedure that allows mean-field theories to be constructed for arbitrary sets of master equations, and as an example consider the case of reflective defects. We derive an expression for the FM chirp of arbitrary Fabry-Perot cavities---important for most realistic lasers---and use perturbation theory to show how they are affected by gain curvature. Lastly, we show that an eigenvalue formulation of the laser's dynamics can be useful for characterizing all of the stable states of the laser: the fundamental comb, the continuous-wave solution, and the harmonic states.
\end{abstract}

\section{Introduction}

Frequency-modulated (FM) combs have long existed in lasers \cite{yarivInternalModulationMultimode1965}, and in particular tend to arise in semiconductor lasers, which can spontaneously enter self-FM regimes \cite{hugiMidinfraredFrequencyComb2012,khurginCoherentFrequencyCombs2014a,burghoffTerahertzLaserFrequency2014,singletonEvidenceLinearChirp2018,hillbrandInPhaseAntiPhaseSynchronization2020,sterczewskiFrequencymodulatedDiodeLaser2020,daySimpleSingleSectionDiode2020,krisoSignaturesFrequencymodulatedComb2021,heckPassivelyModeLocked102009}. There are many applications of laser frequency combs, particularly dual-comb spectroscopy \cite{villaresDualcombSpectroscopyBased2014,yangTerahertzMultiheterodyneSpectroscopy2016} and radiometry  \cite{benirschkeFrequencyCombPtychoscopy2021}. Frequently, these combs exhibit an interesting linearly-chirped behavior \cite{singletonEvidenceLinearChirp2018,hillbrandInPhaseAntiPhaseSynchronization2020,sterczewskiFrequencymodulatedDiodeLaser2020}, which was not theoretically predicted. This behavior was first conclusively identified in quantum cascade lasers (QCLs) \cite{singletonEvidenceLinearChirp2018}, which are notable for their fast gain recovery times, but have also been identified in slow gain recovery time systems as well \cite{sterczewskiFrequencymodulatedDiodeLaser2020,daySimpleSingleSectionDiode2020,krisoSignaturesFrequencymodulatedComb2021,heckPassivelyModeLocked102009}.

Recently, we developed a mean-field theory (MFT) analogous to the Lugiato-Lefever equation that is essential for understanding the universality of this behavior \cite{burghoffUnravelingOriginFrequency2020}. The Lugiato-Lefever equation has been used to describe many nonlinear resonators, especially Kerr combs \cite{coenModelingOctavespanningKerr2013,lugiatoLugiatoLefeverEquation2018,chemboSpatiotemporalLugiatoLefeverFormalism2013b}. The basic idea is that if the changes to the field are small within a single round-trip, mean-field theories allow one to quickly predict the long term behavior of a system without taking small timesteps. The fundamental difficulty with lasers is that most lasers do \textit{not} have small changes within a round trip: they often have large mirror losses that are balanced by gain. Thus, most simulations that had previously been developed relied either on explicit solutions to the Maxwell-Bloch equations \cite{gordonCoherentTransportSemiconductor2009a,wangActiveModelockingMidinfrared2015a,silvestriCoherentMultimodeDynamics2020,wangHarmonicFrequencyCombs2020,tzenovAnalysisOperatingRegimes2017,dongPhysicsFrequencymodulatedComb2018,khurginCoherentFrequencyCombs2014a,henryPseudorandomDynamicsFrequency2017,henryTemporalCharacteristicsQuantum2018,piccardoFrequencyModulatedCombsObey2019} or simpler coupled master equations \cite{opacakTheoryFrequencyModulatedCombs2019,piccardoFrequencyCombsInduced2020}. These methods require the simulation of many timesteps and in general are not analytically tractable. To address this limitation, we showed that the large changes within a laser could be eliminated by normalizing the electric field to a scaling function that essentially preconditions the master equations \cite{burghoffUnravelingOriginFrequency2020,burghoffSupplementaryDocumentFrequencymodulated2020}. This scaling function is found using the steady-state power profile of the laser, i.e., the power produced by a laser without any phase-dependent terms. Note that this normalization is essential for most Fabry-Perot semiconductor lasers (whose typical mirror reflectivities are around 30\%), but it is not needed for ring lasers (whose coupling losses are much smaller). Thus, suitable mean-field theories can be derived for rings by setting this scale factor to unity \cite{columboUnifyingFrequencyCombs2021a}.

Although active cavity mean-field theory can apply to several different systems \cite{burghoffUnravelingOriginFrequency2020,columboUnifyingFrequencyCombs2021a,jaidlCombOperationTerahertz2021}, in this work we focus on the system show in Fig. \ref{fig:Cavity}. This setup describes a two-level laser gain medium inside a Fabry-Perot (FP) cavity. If one then makes some weak simplifying assumptions (a slowly varying envelope, a constant amplitude, and linear internal power growth) the analytical result follows in the form of a phase-dominated nonlinear Schrödinger equation (phase NLSE) \cite{burghoffUnravelingOriginFrequency2020},
\begin{equation} \label{eqn:NLSE}
-i \frac{\partial F}{\partial T} = \frac{\beta}{2} \frac{\partial^2 F}{\partial z^2} + \gamma \left|F\right|^2 \left(\arg F-\langle \arg F \rangle\right) F +i r \left(\left|F\right|^2-P_0\right)F,
\end{equation}
where $F$ is the electric field normalized to the unitless intracavity gain (proportional to the laser’s steady-state intracavity intensity), T is the slow time, $\beta$ is the normalized group velocity dispersion (GVD), $\gamma$ is the nonlinear cross-steepening, angle brackets represent an average over position, and $r$ represents amplitude relaxation to the steady state intensity at the left facet, $P_0$. The position $z$ is taken within the artificially extended laser cavity---if $L_c$ is the physical cavity length, wave traveling in the positive direction are are mapped to $z \in \left[0,L_c\right)$, and waves traveling in the negative direction are mapped to $z \in \left[-L_c,0\right)$. Previously, it was shown that the fundamental solution to this equation can be written analytically as \cite{burghoffUnravelingOriginFrequency2020}
\begin{equation} \label{eqn:fundamental}
F_1\left(z,T\right) =A_0 \exp\left[i \frac{\gamma \left|A_0\right|^2}{2 \beta} \left(z^2-\frac{1}{3}L_c^2 \gamma \left|A_0\right|^2 T\right) \right],
\end{equation}
where $A_0=\sqrt{\frac{P_0}{1+\gamma/2r}}$. This solution was referred to as an extendon due to its spatially-extended nature. The higher-order solutions to the phase NLSE are harmonic states \cite{piccardoFrequencyModulatedCombsObey2019,forrerSelfstartingHarmonicComb2021}, and the $N$th one can be taken as \cite{burghoffUnravelingOriginFrequency2020}
\begin{equation} \label{eqn:hstate}
F_N\left(z,T\right) = F_1\left(z\bmod \frac{2L_c}{N},T\right),
\end{equation}
where the modulo operator is taken over $z\in\left[-\frac{L_c}{N},\frac{L_c}{N}\right)$. Fabry-Perot lasers can sometimes prefer harmonic modes of operation, but even when the fundamental state is preferred the harmonic states detailed in Eq. (\ref{eqn:hstate}) are useful for forming a basis set for further analysis.

In this work, we extend these results to derive several key results pertaining to FM combs. First, we generalize the previous mean-field theory from Ref. \cite{burghoffUnravelingOriginFrequency2020}, showing how it can be applied to arbitrary models of the laser under consideration---two-level models, three-level models, systems with defects, etc. An algorithm is provided describing the steps one needs to follow to convert an arbitrary master equation into its corresponding mean-field theory.

Next, we generalize previous results to allow for arbitrary cavities. In order to derive the phase NLSE, it was previously assumed that one of the end mirrors of the Fabry-Perot cavity have unity reflectivity (R=100\%). It was not shown that the phase NLSE applies to cavities with two non-unity reflectors, which includes most lasers. An analysis is presented here to expand this previous result. This is a useful pursuit because the mirror losses are integral for optimizing the the spectral bandwidth \cite{beiserEngineeringSpectralBandwidth2021a}.

Then, we improve the accuracy of the extendon theory in the face of gain curvature. Though the analytical solution of the NLSE (Eq. (\ref{eqn:NLSE})) matches the numerical solution while neglecting gain curvature (everywhere but the turnaround point), when more realistic gain curvature was applied the accuracy dips noticeably \cite{burghoffUnravelingOriginFrequency2020}. To address this, we use a perturbative analysis to show the relevant parameters can be modified when gain curvature is present. These results allow the numerical and analytical results to match closely in the case of moderate gain curvature, though at very high values the accuracy drops.

Lastly, we show that by calculating the diagonal elements of the matrix corresponding to Eq. (\ref{eqn:NLSE}), one can gain insight into the dynamics of the phase NLSE. These calculations provide the carrier envelope offset (CEO) of the comb---the offset of the frequency comb teeth. This method also provides limited insights into the stability of the various harmonic states. In particular, we show that the addition of reflective defects---which has been shown to be an effective way of modifying harmonic FM combs \cite{kazakovDefectengineeredRingLaser2021}--- adds to the imaginary component of these diagonal elements, which can improve the stability of certain eigenstates while weakening the others. For sufficiently large defect magnitudes, this results in the laser favoring harmonic states rather than fundamental states, or for moderate magnitudes can generate quasi-stable solutions which breath periodically in time.

\section{Analysis}
\medskip

\subsection{Laser mean field theory for arbitrary sets of master equations}
\medskip

\begin{figure}\centering
	\includegraphics[scale=0.5]{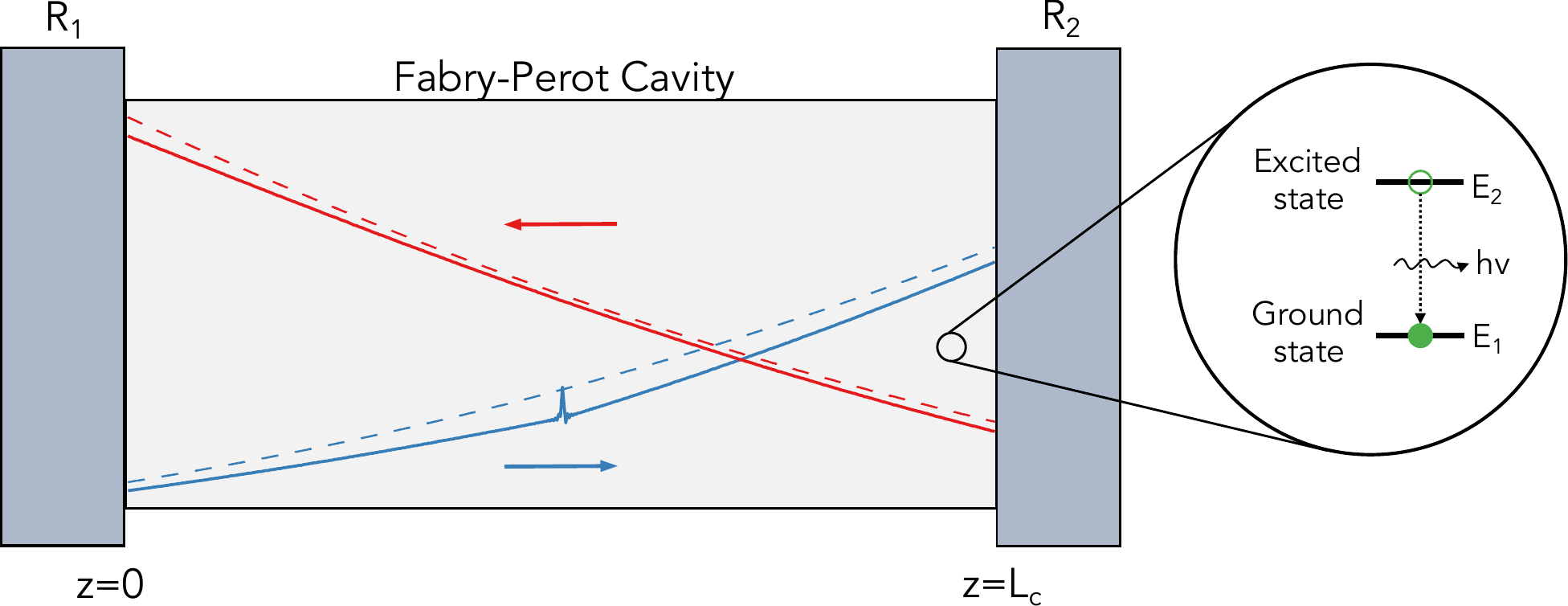}
	\caption{\label{fig:Cavity} Diagram of a Fabry-Perot laser, showing the intensity of waves traveling in each direction as solid lines. The intensity closely mimics the steady-state power (i.e., the continuous wave solution), which are shown here as dashed lines. These quantities are plotted for the case of $R_1=0.2$ and $R_2=0.45$.}
\end{figure}

We wish to expand and generalize the previous work's mean-field theory analysis. The prior theory was derived specifically for the set of laser master equations describing a two-level system, but other formulations are possible depending on the physics one wishes to consider. For example, in section \ref{sec:defect} we will consider the case of adding reflective defects to the cavity. In this section, we will closely examine how one can convert arbitrary master equation terms into appropriate mean-field terms. To begin, consider a two-level system, whose master equations can mostly simply be expressed as
\begin{align}  \label{eqn:master_eqn}
\frac{n}{c} \frac{\partial E_\pm}{\partial t} + \frac{\partial E_\pm}{\partial z} &= \frac{g_0}{2}\left(1-\frac{1}{P_s}\left(\left|E_\pm\right|^2+2\left|E_\mp\right|^2\right) \right) E_\pm \nonumber -\frac{\alpha_w}{2} E_\pm \\ & +i\frac{1}{2}k''\frac{\partial^2 E_\pm}{\partial t^2} + \frac{g_0}{2}T_2^2 \frac{\partial^2 E_\pm}{\partial t^2} \\ 
&+\frac{g_0}{2 P_s}\left(\left(2T_1+3T_2\right) \frac{\partial E_\mp^\ast}{\partial t} E_\mp +\left(T_1+\tfrac{5}{2}T_2\right) E_\mp^\ast \frac{\partial E_\mp}{\partial t} \right)E_\pm, \nonumber
\end{align}
where $E_\pm$ is the envelope of the wave traveling in the $\pm z$ direction (see Fig. \ref{fig:Cavity} for reference), $g_0$ is the small signal gain, $k''$ is the group velocity dispersion (GVD), $T_1$ and $T_2$ are the population and coherence lifetimes for a two-level system, and $n$ is the refractive index of the medium. The first line represents losses, gain, and gain saturation, the second represents dispersion and gain curvature, and the final lines cross-steepening. As previously discussed, the most critical terms for FM comb generation are the cross-steepening terms, which arise due to temporal variation of the gain grating. Other effects can be included as well, but these serve only to alter the parameters shown here. (For example, Kerr nonlinearity and linewidth enhancement shift the effective dispersion \cite{burghoffUnravelingOriginFrequency2020}.) Alternatively, the master equations of a three level system can be described by \cite{beiserEngineeringSpectralBandwidth2021a}
\begin{equation} \label{eqn:master_eqn_Vienna}
    \begin{aligned} 
    \left(\frac{n}{c} \partial_t + \partial_z \right)E_\pm &= \frac{g}{2} \frac{1+ib}{1+i\xi} \left[1-\tilde{T_2}+\tilde{T_2}^2 \partial_t^2 \right]E_\pm\\
    &-\frac{g T_g}{T_1 I_\textrm{sat}} \frac{1+ib}{1+i\xi}\left[|E_\mp|^2 E_\pm - \left(\tilde{T_2}+T_g\right)|E_\mp|^2 \partial_t E_\pm \right.\\ 
    & \left. -\left(\tilde{T_2}+T_1\right)E_\pm E_\mp \partial_t E_\mp^*-\tilde{T_2} E_\pm E_\mp^* \partial_t E_\mp\right]-\frac{\alpha_w}{2}E_\pm,
    \end{aligned}
\end{equation}
where $g = \frac{g_0}{1+\frac{I}{I_sat}}$ is the saturated gain, $I_\textrm{sat}$ is saturated intensity, and $I=|E_-|^2+|E_+|^2$ the normalized intensity. 

In order to unify these descriptions, note that Equation \ref{eqn:master_eqn} can be written more generally as
\begin{equation} \label{eqn:MasterGeneral}
    \frac{n}{c}\frac{\partial{E_\pm}}{\partial{t}}+\frac{\partial E_\pm}{\partial z}=\sum_i 
    p_i(z)\widehat{Q}_i(E_\pm(z,t))\widehat{R}_i(E_\mp(z,t)),
\end{equation}
where $p_i(z)$ is a time invariant function of position and $\widehat{Q}_i$ and $\widehat{R}_i$ are operators acting on $E_\pm$ and $E_\mp$, respectively. This is valid because every term in the right-hand side of (\ref{eqn:master_eqn}) is separable. Though (\ref{eqn:master_eqn_Vienna}) is not separable, it can be written in the same way provided that one uses a series expansion to take $g={}g_0\left(1-\frac{I}{I_\textrm{sat}} +...\right)$. This is the general form we will therefore consider.

Next, backward propagating waves are flipped and extended to $\left(-L_c,0\right)$, where $L_c$ is the cavity length. This artificial cavity extension reduces the master equations to a single forward propagating wave on $\left(-L_c, L_c\right)$, and periodic boundary conditions now apply. Mathematically, this corresponds to making the substitutions $E_\pm(z,t) \rightarrow E(z,t)$ and $E_\mp(z,t) \rightarrow E(-z,t)$. For convenience, we adopt the notation $E_-(z)\equiv E(-z)$. The only modification needed to the master equation (aside from changing the domain of the fields) is to replace explicit space dependence ($p_i(z)$) with an appropriately modified piecewise-defined function, $\tilde{p}_i(z)$. After this, the steady-state intensity $P\left(z\right)$ of the system with no phase dynamics is then calculated, taking into account gain saturation, waveguide loss, and mirror losses.
This intensity is typically very similar to the master equation solutions (see Fig. \ref{fig:Cavity}), and so normalizing by it eliminates large changes due to the facets. Mathematically, we define the slow-intensity envelope $F(z,t)$ using $E\left(z,t\right)\equiv K^{1/2}\left(z\right) F\left(z,t\right)$. Here $K\left(z\right)$ can be thought of as the dimensionless power gain that the wave experiences as it traverses the cavity. It is time-independent, and for a laser above threshold it is related to the steady-state intensity by $K\left(z\right)=P\left(z\right)/P_0$, where $P_0$ is the power at the left cavity facet. It can alternatively be expressed in terms of the effective gain as
\begin{equation} \label{eqn:K_EffGainExp}
    K(z) = \exp\left({\int_0^{z} g_\textrm{eff}(z')dz'}\right),
\end{equation}
where $g_\textrm{eff}$ contains the effects of gain saturation, waveguiding loss, and mirror losses. 
This change of variable further simplifies the problem, in particular removing delta function mirror losses. Performing this substitution, one finds that the general master equation can now be rewritten as

\begin{align}
    \frac{n}{c}\frac{\partial{F}}{\partial{t}}+\frac{\partial F}{\partial z}&=\sum_{i} 
    \frac{\tilde{p}_i(z)}{\sqrt{K(z)}}\,\widehat{Q}_i\left(\sqrt{K(z)}F(z,t)\right)\widehat{R}_i\left(\sqrt{K_-(z)}F_-(z,t)\right) \nonumber \\
    &- \sum_{i\in P} 
    \frac{\tilde{p}_i(z)}{\sqrt{K(z)P_0}}\,\widehat{Q}_i\left(\sqrt{K(z) P_0}\right)\widehat{R}_i\left(\sqrt{K_-(z)P_0}\right)F(z,t)  \label{eqn:PQR}
\end{align}
where the first line contains all terms from the master equation and the second contains just those terms contained within the power calculation. Most terms included in the power calculation will now vanish. For example, for (\ref{eqn:master_eqn}) one obtains
\begin{align} \label{eqn:fbeqn} 
\frac{n}{c} \frac{\partial F}{\partial t} + \frac{\partial F}{\partial z} =  
&-\frac{g_0}{2P_s}\left(K \left(\left|F\right|^2-P_0\right)+2K_-\left(\left|F_-\right|^2 - P_0\right) \right) F 
+i\frac{1}{2}k''\frac{\partial^2 F}{\partial t^2} + \frac{g_0}{2}T_2^2 \frac{\partial^2 F}{\partial t^2} \nonumber \\
&+\frac{g_0}{2 P_s}K_-\left((2T_1+3T_2) \frac{\partial F_-^\ast}{\partial t} F_- +(T_1+\tfrac{5}{2}T_2) F_-^\ast \frac{\partial F_-}{\partial t} \right)F.
\end{align}
The mirror losses, waveguide losses, and small-signal gain have vanished; the gain saturation is now driving the system towards a power of $P_0$. Again, (\ref{eqn:PQR}) can be compactly written as 
\begin{equation} \label{eqn:NormalGeneral}
    \frac{n}{c}\frac{\partial{F}}{\partial{t}}+\frac{\partial F}{\partial z}=\sum_i 
    f_i(z)\widehat{G}_i(F(z,t))\widehat{H}(F_-(z,t)),
\end{equation}
 where $\widehat{G}_i$ and $\widehat{H}_i$ are operators acting on $F$ and $F_-$, respectively, and $f_i(z)$ is a time invariant function of position.
\begin{figure}\centering
	\includegraphics[scale=0.5]{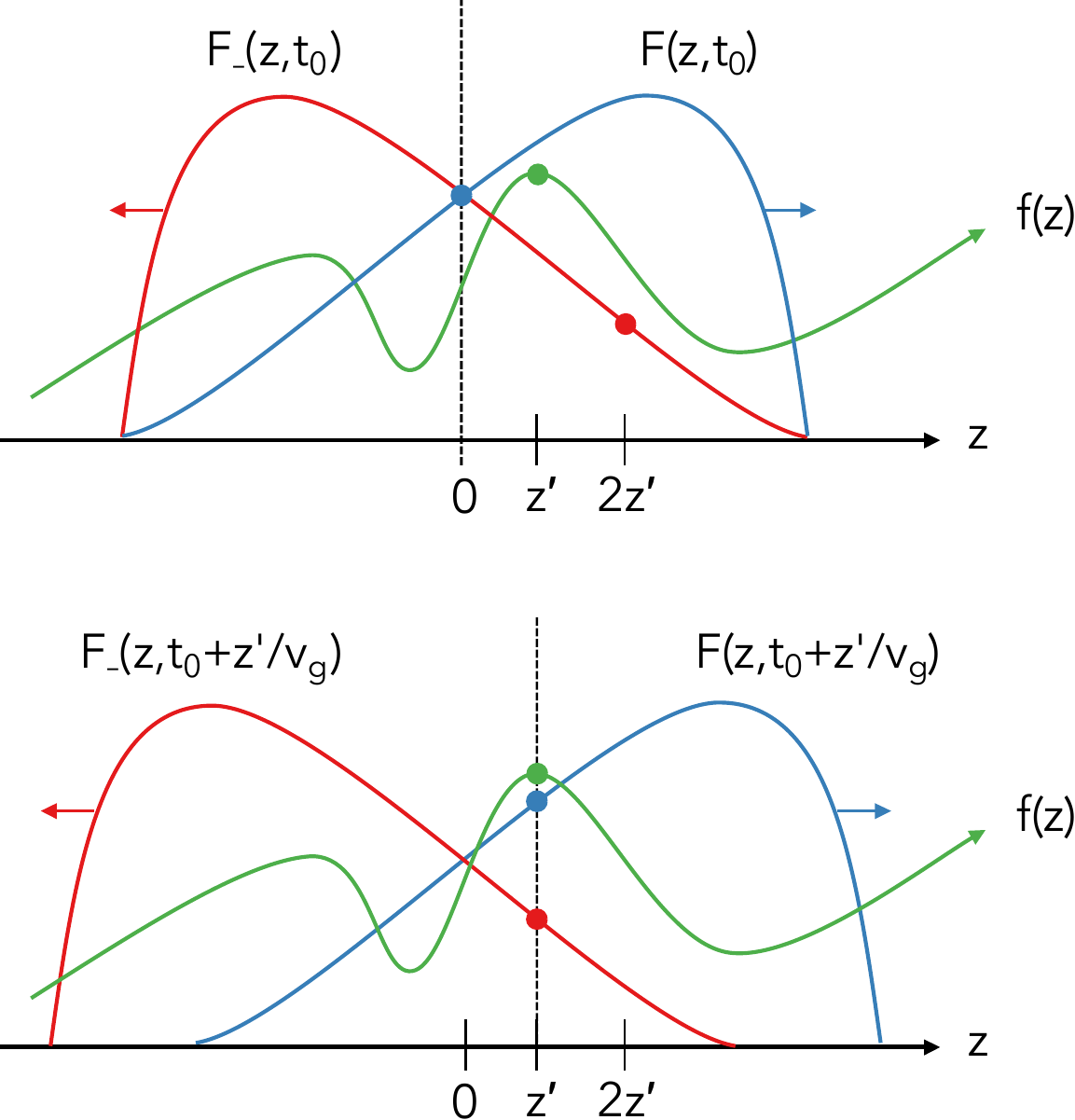}
	\caption{\label{fig:FB}How the forward traveling envelope ($F$) changes relative to the backward traveling envelope ($F_-$) and an arbitrary fixed time function ($f$). Note the origin is fixed to a point on the forward traveling wave; that is, the origin moves with the forward traveling wave (as indicated by the dashed line).}
\end{figure}

At this point, one must integrate Eq. (\ref{eqn:NormalGeneral}) over a full round trip over the cavity to finish the mean field derivation. The left-hand side of Eq. (\ref{eqn:NormalGeneral}) is the typical wave equation and is periodic, so it vanishes with averaging. Hence, the only relevant round trip change comes from the sum of the right-hand side of the equation over every point during a round trip of the cavity. Therefore, (\ref{eqn:NormalGeneral}) becomes 
\begin{equation} \label{eqn:MFTint}
    \begin{aligned}
    \frac{\partial F}{\partial T} = \frac{\Delta F}{T_r} =\frac{c}{2L_cn}\sum_i \int_{-L_c}^{L_c} f_i\left(z+z'\right)\widehat{G}_i\left(F\left(z\right)\right)\widehat{H}_i\left(F\left(-z-2z'\right)\right)dz',
    \end{aligned}
\end{equation}
where $T_r$ is the round trip time (the total time it takes for the wave to travel the extended cavity), $T$ is a slow time, usually described in fractional units of $T_r$, and $z\in\left(-L_c,L_c\right)$. To understand why each function is offset differently in Eq. (\ref{eqn:MFTint}), one should consult Fig. \ref{fig:FB}. For this averaging, assume the origin is attached to a point on the forward traveling envelope ($F$). After a short time of $z'/v_g$ passes, stationary functions such as $f\left(z\right)$ shift to left $\left(f\left(z\right)\rightarrow f\left(z+z'\right)\right)$ and backward traveling envelopes change according to $F_-\left(z\right)=F\left(-z\right)\rightarrow F\left(-z-2z'\right)=F_-\left(z+2z'\right)$. The second case involving backward envelopes is not quite as intuitive as the first, but if one considers that the forward envelope (origin) moves to the right z’ and the backward envelope moves to the left $z’$, it is apparent the overall shift is $2z’$. Armed with this information, one just needs to form the convolution integral to deal with backward propagating waves, and it becomes straightforward to apply the generalized mean-field theory. Defining this integral as
\begin{equation} \label{eqn:Kconv_general}
    \widetilde{f}\left[q\right](z)=\frac{1}{2L_c} \int_{0}^{2L_c}f\left(-u\right) q\left(z-2u\right)du.
\end{equation}
With Eq. (\ref{eqn:Kconv_general}), the final result can be written compactly as
\begin{equation} \label{eqn:MFT_result}
    \begin{aligned}
    \frac{n}{c} \frac{\partial F}{\partial T}=\sum_i \widehat{G}_i\left(F\right)\widetilde{f}[\widehat{H}_i\left(F\right)].
    \end{aligned}
\end{equation}
For convenience, many common terms converted from master equations to mean-field theory are detailed in Table (\ref{tab:MFTalg}). If desired, the spatiotemporal substitution \cite{chemboSpatiotemporalLugiatoLefeverFormalism2013b} $\partial_t \rightarrow -\frac{c}{n} \partial_z$ can be used to convert fast-time derivatives to spatial derivatives, or vice versa.

\renewcommand{\arraystretch}{2}
\arrayrulecolor{black}
\begin{center}
\begin{table}[!h] 
\centering
\begin{tabular}{||c|c|c||} 
\hline
 {\bfseries Master equation} & {\bfseries Normalized extended cavity} & {\bfseries Mean-field theory} \\  
 \hline
 $p\,\widehat{Q}(E_\pm)\widehat{R}(E_\mp)$ & $f\,\widehat{G}\left(F\right)\widehat{H}\left(F_-\right)$ & $ \widehat{G}\left(F\right) \widetilde{f}\left[\widehat{H}\left(F\right)\right]$\\
 \hline
 \cellcolor{gray!25}$\tfrac{1}{2}\left(-\alpha_w+g_0\right)E_\pm$ 
    & \cellcolor{gray!25}0 
    & \cellcolor{gray!25}0 \\ \hline
 $ \cellcolor{gray!25}-\frac{g_0}{2P_s}\left(|E_\pm|^2+2|E_\mp|^2)\right)E_\pm$
    & \cellcolor{gray!25}\makecell{\small $-\frac{g_0}{2P_s}\left(K \left(\left|F\right|^2-P_0\right)\right.$ \\
       \small $\left.+2K_-\left(\left|F_-\right|^2 - P_0\right) \right) F$}
    & \cellcolor{gray!25} \makecell{\small $-\frac{g_0}{2P_s}\left(\langle K \rangle \left(\left|F\right|^2-P_0\right)\right.$ \\
      \small $\left.+2\left(\widetilde{K}\left[\left|F\right|^2\right] - \langle K \rangle P_0\right) \right) F$}\\
 \hline
 $E_\pm$ & $F$ & $F$ \\ 
 \hline
$E_\mp$ & $\frac{K_-^{1/2}}{K^{1/2}}F_-\equiv K_\mathrm{frac} F_-$ & $ \widetilde{K}_\mathrm{frac}\left[F\right]$ \\ 
 \hline
$|E_\mp|^2 \frac{\partial E_\pm}{\partial t}$ & $ K_- |F_-|^2 \frac{\partial F}{\partial t}$ & $\widetilde{K}\left[|F|^2\right] \frac{\partial F}{\partial t}$ \\
 \hline
$E_\mp \frac{\partial E_\mp}{\partial t}^* E_\pm $
    & $ K_- F_- \frac{\partial F_-}{\partial t}^* F $ 
    & $ \widetilde{K} \left[F \frac{\partial F}{\partial t}^*\right] F$ \\
 \hline
 $|E_\pm|^2 \frac{\partial E_\pm}{\partial t}$ & $ K |F|^2 \frac{\partial F}{\partial t}$ & $\langle K \rangle |F|^2 \frac{\partial F}{\partial t} $ \\
 \hline
 $E_\pm \frac{\partial E_\pm}{\partial t}^* E_\pm$ & $ K F \frac{\partial F}{\partial t}^* F$ & $ \langle K \rangle F \frac{\partial F}{\partial t}^* F$ \\
 \hline
\end{tabular}

\caption{Conversion of common master equation terms to normalized extended cavity and mean-field theory. Terms in gray are associated with the steady-state power calculation and mostly vanish. $\widetilde{f}[q](z)$ is defined according to Eq. (\ref{eqn:Kconv_general}) and $\langle f(z) \rangle = \frac{1}{2L_c}\int_0^{2L_c} f(z) dz$ is an average over the cavity space.} 
\label{tab:MFTalg}
\end{table}
\end{center}

For the rest of this work, we rely on the two-level system---for reference, this leads to an evolution equation of the form
\begin{align}  \label{eqn:twolevelmeanref}
\frac{n}{c} \frac{\partial F}{\partial T} = &\left(\frac{c}{n}\right)^2\left(i\frac{1}{2}k'' + \frac{g_0}{2}T_2^2\right) \frac{\partial^2 F}{\partial z^2} 
-\frac{g_0}{2P_s}\left(\langle K \rangle \left(\left|F\right|^2-P_0\right)\right.
      \left.+2\left(\widetilde{K}\left[\left|F\right|^2\right] - \langle K \rangle P_0\right) \right) F
\nonumber \\
&-\left.\frac{c}{n}\right.\frac{g_0}{2 P_s}\left((2T_1+3T_2) \widetilde{K} \left[F \frac{\partial F}{\partial z}^*\right] + (T_1+\tfrac{5}{2}T_2) \widetilde{K} \left[F^* \frac{\partial F}{\partial z}\right] \right)F.
\end{align}
In Appendix \ref{sec:nonlingv}, the effect of additional terms are investigated, including additional third-order nonlinearities and first-order linewidth terms. In general, these do not have major effects. Primarily, they either modify extendon formation by a few percent or slightly adjust its group velocity.


\subsection{Phase potential NLSE of a variable reflectivity cavity}

In previous work, it was shown that a Fabry-Perot cavity with one mirror of unity reflectivity obeyed a nonlinear Schrodinger equation with a phase potential, but this result did not generalize to arbitrary cavities. From numerical simulations it is apparently the case that the result actually should generalize, but it was not proven and no expression for the important chirp bandwidth was derived for the case where $R_1$ and $R_2$ is allowed to be variable. We perform this analysis here. While the details are more complicated, the final result is similar to the unity cavity result. Consider the cross-steepening terms of the two-level system mean-field theory, which ultimately give rise to a phase potential. They take the form
\begin{equation}\label{eqn:power_terms}
    \begin{aligned}
    \left(\frac{\partial F}{\partial T}\right)_\mathrm{XS} &= -\left(\frac{c}{n}\right)^2\frac{g_0}{2P_s}\left( \left(2T_1+3T_2\right) \tilde{K}\left[ F\tfrac{\partial F^\ast}{\partial z}\right] +\left(T_1+\tfrac{5}{2}T_2\right) \tilde{K}\left[F^\ast \tfrac{\partial F}{\partial z}\right]  \right) F.
    \end{aligned}
\end{equation}
In order to evaluate this for a general cavity, two assumptions are made. First, the averaged field envelope ($F$) is assumed to have a constant amplitude ($A$). This is typically a very good approximation, as fast gain saturation suppresses amplitude fluctuations. Doing so allows the amplitude to factor out of the convolution and leaves only the phase ($\phi$), since $\tilde{K}\left[F\tfrac{\partial F^\ast}{\partial z} \right]=-i|A|^2\tilde{K}\left[\tfrac{\partial \phi}{\partial z}\right]=-\tilde{K}\left[F^\ast \tfrac{\partial F}{\partial z}\right]$. Secondly, the steady-state power profile is assumed to be piecewise linear (see Figure \ref{fig:PowerProf}). Both facets introduce a discontinuity that will ultimately give rise to a delta function when (\ref{eqn:power_terms}) is evaluated. The system's power profile and its derivative can be simply expressed as follows:

\begin{figure}\centering
	\includegraphics[scale=0.45]{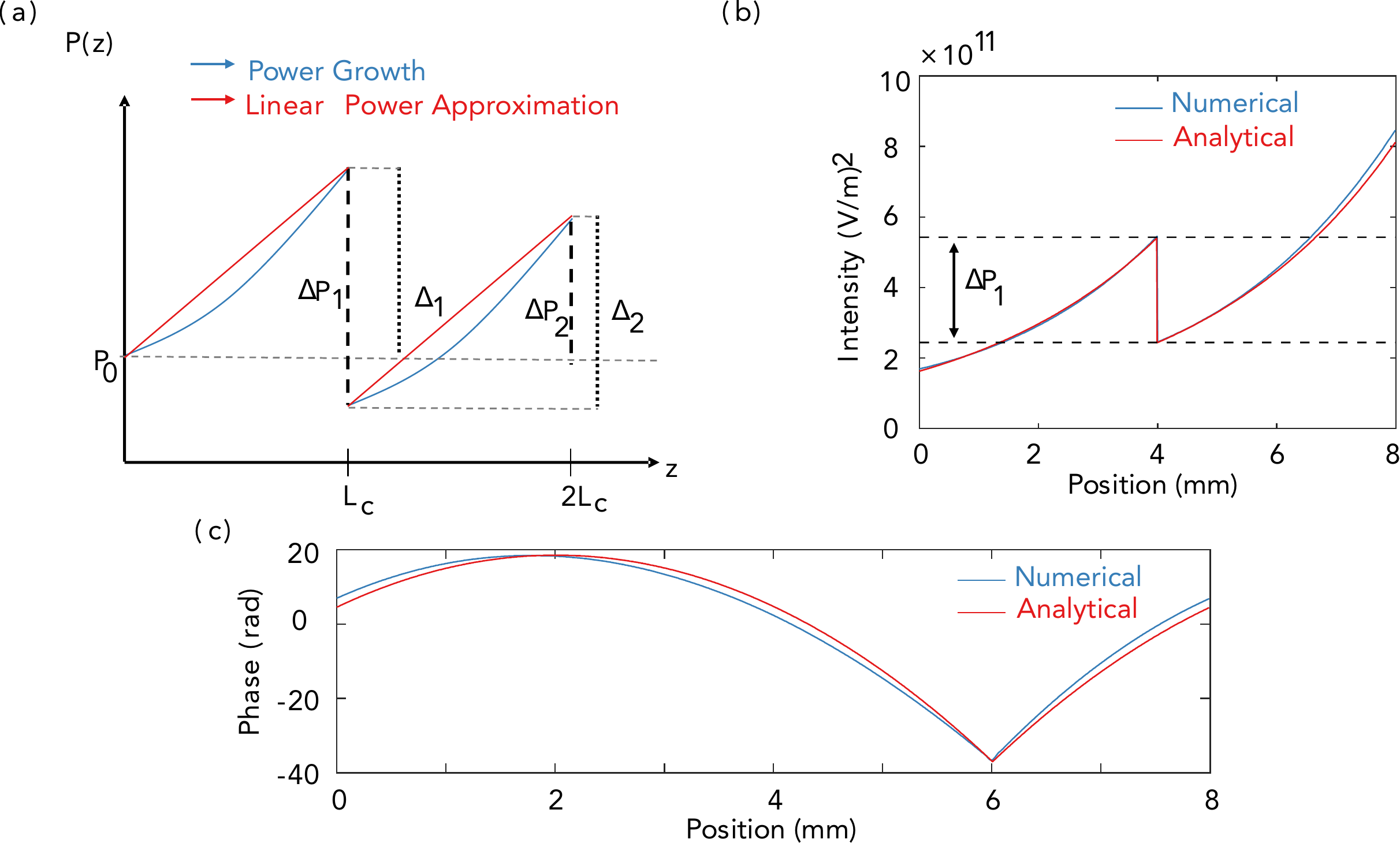}
	\caption{\label{fig:PowerProf}(a) Power profile sketch showing key features like initial power at the left facet, $P_0$ , and the corresponding power drops at the right facet ($\Delta P_1$) and the left facet ($\Delta P_2$). (b) The analytical power function, Eq. (\ref{eqn:power_actual}), versus the numerical power function. Here $R_1=0.2$ and $R_2=0.45$. Note that the values for $\Delta P_1$ are from the analytical expression Eq. (\ref{eqn:drop1}). Note that the data in this plot has not been normalized by $K(z)$. (c) Numerical and analytical phase function calculated using (\ref{eqn:chirpBW}), showing good agreement. Parameter values are included in Table \ref{tab:SimValues}.}
\end{figure}

\begin{equation}\label{eqn:steadypower}
  P\left(z\right) =
  \begin{cases}
    P_0+\frac{z}{L_c} \Delta_1, & \text{for } 0\le z <L_c \\
    \left(1-R_2\right)\left(P_0+\frac{1}{L_c} \Delta_1\right)+\frac{z-L_c}{L_c} \Delta_2, & \text{for } L_c\le z<2L_c
  \end{cases}
\end{equation}
and
\begin{equation}\label{eqn:steadypower_der}
  \frac{\partial P}{\partial z} =
  \begin{cases}
    \frac{\Delta_1}{L_c} -\Delta P_2\delta\left(z\right), & \text{for } 0\le z <L_c \\
    \frac{\Delta_2}{L_c} -\Delta P_1\delta\left(z-L_c\right), & \text{for } L_c\le z<2L_c
  \end{cases}
\end{equation}
Within the constant amplitude assumption, the only terms that can change drastically with a general cavity are those containing a $\widetilde{K}\left[\partial\phi/\partial z\right]$. This is because it is the main operator that relies on the steady-state power explicitly---the steady-state power also affects $\langle K \rangle$ but this quantity is very straightforward to calculate. To evaluate $ \widetilde{K}\left[\partial\phi/\partial z\right]$, one starts with (\ref{eqn:Kconv_general}) to obtain
\begin{align} \label{eqn:K_conv_dphidz}
    \widetilde{K}\left[\frac{\partial\phi}{\partial z}\right]&=\frac{1}{2L_c} \int_{z}^{z+2L_c}K\left(-u\right) \frac{\partial\phi}{\partial z}\left(z-2u\right)du \\
    &=\frac{1}{2L_c} \left[K\left(-u\right)\phi\left(z-2u\right)\right]_0^{2L_c}-\frac{1}{2L_c}\int_{z}^{z+2L_c}\frac{\partial K}{\partial z}\left(-u\right)\phi\left(z-2u\right)du.
\end{align}
Now, recalling that $ K\left(z\right)=P\left(z\right)/P_0\ $ and that the first term goes to zero (because $K\left(z\right)$ and $ \phi\left(z\right)$ are both periodic with respect to $2L_c$), one obtains the following:
\begin{align}\label{eqn:K_conv_work}
    \widetilde{K}\left[\frac{\partial\phi}{\partial z}\right]={} & -\frac{1}{2L_c P_0} \int_{z}^{z+2L_c}\frac{\partial P}{\partial z}\phi\left(z-2u\right)du\nonumber\\
    & =-\frac{1}{2L_c P_0} \left[\int_{0}^{L_c}\left(\frac{\Delta_1}{L_c} -\Delta P_2 \delta\left(z\right)\right)\phi\left(z-2u\right)du \right. \nonumber  \\
    & \hspace{1.5cm}\left. +\int_{L_c}^{2L_c}\left(\frac{\Delta_2}{L_c} - \Delta P_1 \delta\left(z-L_c \right)\right)\phi\left(z-2u\right)du\right]\nonumber\\
    & =-\frac{1}{2L_c P_0} \left[\left(\Delta_1+\Delta_2\right)\langle \phi \rangle -\left(\Delta P_1+\Delta P_2\right)\phi\left(z\right)\right]\nonumber
\end{align}
Now, observe that as they are defined in Fig. \ref{fig:PowerProf}, it is the case that $\Delta_1+\Delta_2=\Delta P_1+\Delta P_2\equiv\Delta P.$ Therefore, the final result is that
\begin{equation}\label{eqn:K_conv_fin}
    \widetilde{K}\left[\frac{\partial\phi}{\partial z}\right]=\frac{1}{2L_c} \frac{\Delta P}{P_0} \left(\phi\left(z\right)-\langle \phi \rangle\right).
\end{equation}
This is the same result obtained in Ref. \cite{burghoffUnravelingOriginFrequency2020} for a cavity with a unity reflector and leads to the same NLSE, the only difference being that there are two $ \Delta P $ components. The mean-field result then takes the form
\begin{equation} \label{eqn:NLSE2}
-i \frac{\partial F}{\partial T} = \frac{1}{2}\left(\beta-i\frac{D_g}{2}\right) \frac{\partial^2 F}{\partial z^2} + \gamma \left|F\right|^2 \left(\arg F-\langle \arg F \rangle\right) F +i r \left(\left|F\right|^2-P_0\right)F,
\end{equation}
where the normalized parameters are defined in Table \ref{tab:SimParametersPhys} for reference.

However, there is a difference once $ \Delta P $ is calculated in terms of physical parameters. To understand this result, one must apply boundary conditions to obtain expressions for $ \Delta P_1 $ and $ \Delta P_2 $. For these derivations, one assumes the intracavity power profile is exponential and has gain balanced by mirror losses $\alpha_m\equiv\frac{1}{2L_c}\ln\left(\frac{1}{R_1R_2}\right)$ (a valid approximation near threshold):
\begin{equation}\label{eqn:power_actual}
  P\left(z\right) =
  \begin{cases}
    P_0 e^{\alpha_m z}, & \text{for } 0\le z <L_c \\
    R_2 P_0 e^{\alpha_m z}, & \text{for } L_c\le z<2L_c,
  \end{cases}
\end{equation}
With this assumption, it is straightforward to show the expressions for $ \Delta P_1 $ and $ \Delta P_2 $ are:
\begin{equation} \label{eqn:drop1}
    \Delta P_1=\frac{P_0}{\sqrt{R_1R_2}}\left(1-R_2\right) \hspace{0.4cm}\textrm{and}\hspace{0.4cm}
    \Delta P_2=P_0\left(\frac{1}{R_1} -1\right).
\end{equation}
Note that if $ R_2=1$, then $ \Delta P_1=0 $ and $ \Delta P=\Delta P_2 $ as before.

\renewcommand{\arraystretch}{2}
\arrayrulecolor{black}
\begin{center}
\begin{table}[!h] 
\centering
\begin{tabular}{|c c c|} 
 \hline
 {\bfseries Name} & {\bfseries Symbol} & {\bfseries Expression} \\  
 \hline
Dipole matrix element
    & $|\mu|^2$ 
    & $\frac{z_0^2 e J T_1}{L_\textrm{mod} w_0}$ \\
Small signal gain 
    & $g_0$
    & $\frac{N_\textrm{d} z_0^2 \omega_0 J T_1 T_2}{n\epsilon_0 c \hbar L_\textrm{mod}}$ \\ 
Saturation intensity
    & $P_s$ 
    & $\left(\frac{4 T_1 T_2}{\hbar^2}|\mu|^2\right)^{-1}$ \\
Total intensity drop 
    & $\Delta P$ 
    & $\frac{2L_c\alpha_m}{3}P_s\left(1-\frac{1}{g_0}\left(\alpha_w+\alpha_m\right)\right)$ \\
Average intensity 
    & $\langle P \rangle$ 
    & $\frac{\Delta P}{2L_c \alpha_m}$ \\
Left facet intensity 
    & $P_0$ 
    & $\frac{1}{3}\frac{2L_c\alpha_mP_s}{\frac{1}{\sqrt{R_1R_2}}\left(1-R_2\right)+\left(\frac{1}{R_1}-1\right)} \left(1-\frac{1}{g_0}\left(\alpha_w+\alpha_m\right)\right) $ \\ 
\makecell{Normalized group \\velocity dispersion} 
    & $\beta$ 
    & $\left(\frac{c}{n}\right)^3k''$ \\
Nonlinear cross-steepening 
    & $\gamma$ 
    & $\frac{g_0}{2P_{\textrm{s}}}\left(\frac{c}{n}\right)^2\left(T_1+\frac{1}{2}T_2\right)\frac{1}{4L_c}\frac{\Delta P}{P_0}$ \\
Amplitude relaxation 
    & $r$ 
    & $\frac{g_0}{2P_{\textrm{s}}}\frac{c}{n}\frac{3\langle P \rangle}{P_0}$ \\
Gain curvature 
    & $D_g$ 
    & $2g_0T_2^2\left(\frac{c}{n}\right)^3$ \\
Chirp bandwidth
    & $BW$ 
    & $\frac{1}{24\pi} 2L_c \alpha_m \left(g_0-\alpha_w-\alpha_m\right) \frac{1}{k^{\prime\prime}} \left(T_1+\frac{1}{2}T_2\right)$ \\
Extendon amplitude
    & $A_0$
    & $\sqrt{\frac{P_0}{1+\tfrac{\gamma}{2r}}}$ \\

\makecell{Fundamental eigenvalue}
    & $\lambda$
    & $\frac{\gamma \left|A_0\right|^2}{2 \beta} \left(\frac{1}{3}L_c^2 \gamma \left|A_0\right|^2 \right)$ \\
 \hline
\end{tabular}

\caption{Properties of fundamental extendons in terms of physical laser parameters for a general cavity. Note $N_\textrm{d}$ is the atomic density, the dipole matrix element is for a two level laser system, and $w_0$ is the equilibrium population inversion. For numerical values refer to Table (\ref{tab:SimValues}) in Appendix A.}
\label{tab:SimParametersPhys}
\end{table}
\end{center}

\noindent In spite of this similarity, the initial power required at the left facet does change depending on the setup. To derive this result, note that in steady state the gain must approximately equal the loss,
\begin{equation}\label{eqn:gainloss_bal}
    g_0 \left(1-\frac{1}{P_s} \left(P_++2P_-\right)\right)=\alpha_w+\alpha_m.
\end{equation}
Averaging both sides so that $ \left(P_++2P_-\ \right)=3 \langle P \rangle$, the result for $ \langle P \rangle/P_s\  $ is
\begin{equation}\label{eqn:gainloss_balx}
    \frac{\langle P \rangle}{P_s}=\frac{1}{3}\left(1-\frac{1}{g_0} \left(\alpha_w+\alpha_m\right)\right).
\end{equation}
Alternatively, $ \langle P \rangle$ can be calculated as
\begin{align}\label{eqn:avgpower}
    \langle P \rangle={} & \frac{1}{2L_c}\int_{0}^{2L_c} P\left(z\right)dz =\frac{1}{2L_c} \left[\int_{0}^{L_c} P_0 e^{\alpha_m z} dz+\int_{L_c}^{2L_c}R_1 P_0 e^{\alpha_m z} dz\right]
    =\frac{\Delta P}{2L_c \alpha_m}
\end{align}
Combining Eq. (\ref{eqn:gainloss_balx}) and Eq. (\ref{eqn:avgpower}), one gets
\begin{equation}\label{eqn:powerdrop}
    \Delta P=\frac{2L_c\alpha_m}{3}P_s\left(1-\frac{1}{g_0}\left(\alpha_w+\alpha_m\right)\right).
\end{equation}
Finally, using (\ref{eqn:drop1}), one can rewrite (\ref{eqn:powerdrop}) in terms of $ P_0 $ as
\begin{equation}\label{eqn:power1_init}
    P_0=\frac{1}{3}\frac{2L_c\alpha_mP_s}{\frac{1}{\sqrt{R_1R_2}}\left(1-R_2\right)+\left(\frac{1}{R_1}-1\right)} \left(1-\frac{1}{g_0}\left(\alpha_w+\alpha_m\right)\right).
\end{equation}
Hence, an expression for the initial steady-state intensity at the left facet was derived in terms of physical parameters. Due to our formalism, Eq. (\ref{eqn:power1_init}) is not symmetric with respect to $R_1$ and $R_2$ --- it matters which side one takes as a starting point.

Because the form of the results are the same, in the absence of gain curvature the chirp bandwidth of an extendon can be calculated from (\ref{eqn:NLSE}) as $BW=\frac{1}{2\pi}\frac{c}{n}A_0^2 \frac{\gamma}{\beta}2L_c$. By assuming that $A_0^2\approx P_0$ and plugging in (\ref{eqn:powerdrop}) and (\ref{eqn:power1_init})), one can show that the chirp bandwidth formula is unchanged from previous work, resulting in
\begin{equation} \label{eqn:chirpBW}
    BW \approx \frac{1}{24\pi} 2L_c \alpha_m \left(g_0-\alpha_w-\alpha_m\right) \frac{1}{k^{\prime\prime}} \left(T_1+\frac{1}{2}T_2\right).
\end{equation}
For ease of reference, all of these results are tabulated in Table \ref{tab:SimParametersPhys}.

\subsection{Harmonic states: the higher-order eigenstates of the phase NLSE}\label{sec:eigenstates}
\begin{figure}\centering
	\includegraphics[scale=0.65]{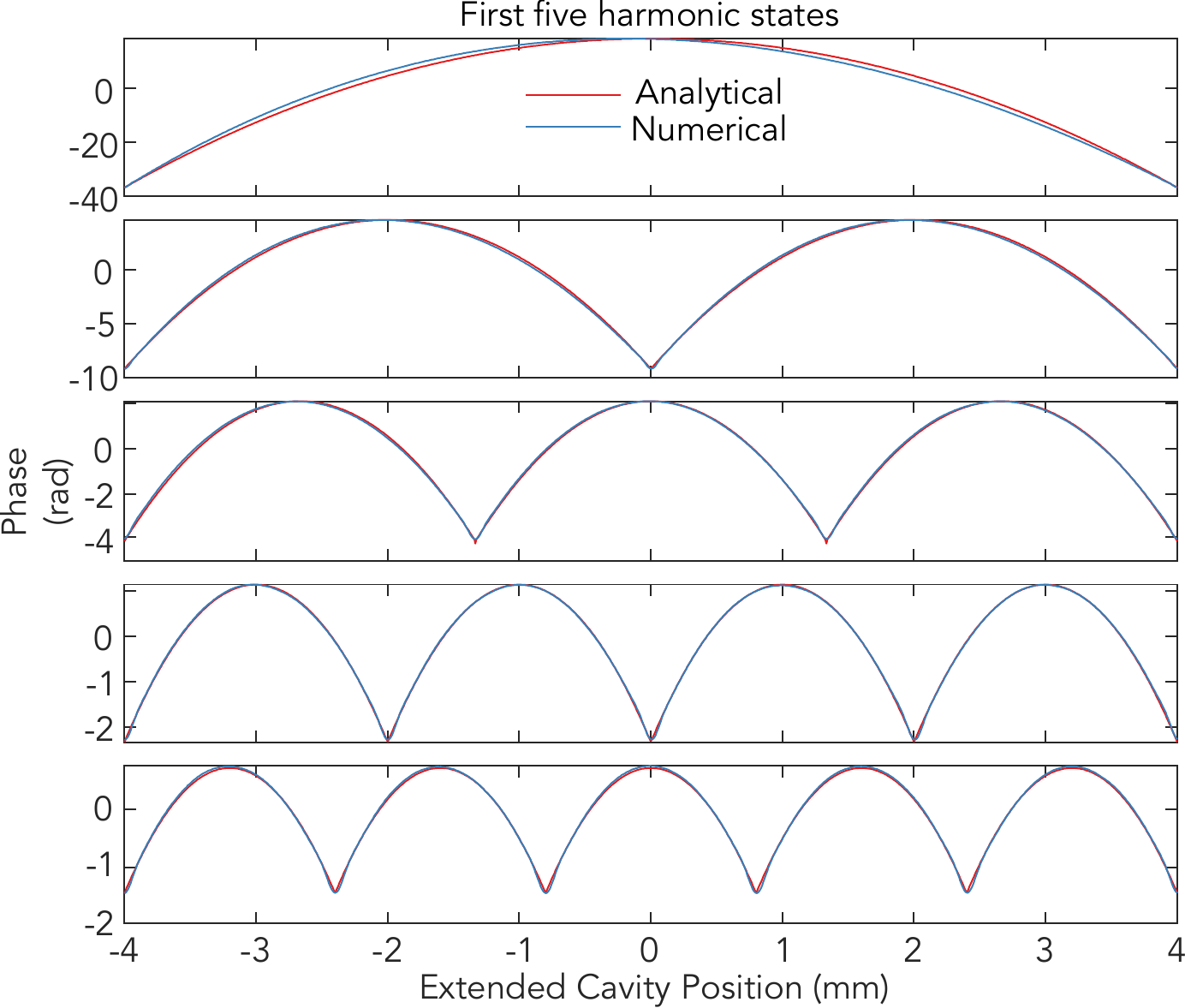}
	\caption{\label{fig:Harmonic}The first five harmonic states plotted in descending order. The analytical (red) data was obtained using Eq. (\ref{eqn:hstate}) and the numerical (blue) data was obtained by numerically solving the mean-field theory. These phases have been arbitrarily offset by their own average for ease of comparison. The numerical harmonic states use gain curvature whereas the analytical harmonic states do not.}
\end{figure}

An interesting consequence of the mean-field theory is that it can reproduce both fundamental FM combs and harmonic combs. By initializing the electric field with a field that is periodic over 1/Nth a round trip, stable harmonic combs are produced rather than fundamental combs. These turn out to be numerically unstable---eventually decaying to the fundamental due to machine precision-level asymmetry---but can be resymmetrized to remain stable indefinitely. The phase profiles of the first five harmonic states are plotted in Figure \ref{fig:Harmonic}. Note that they agree very well with the form predicted by Equation \ref{eqn:hstate}, which constructs them in terms of the fundamental.

In order to better understand the properties of these states, the phase NLSE can be reformulated as an eigenvalue problem expressed in a finite basis of the harmonic extendons. By computing matrix elements and applying the well-known LCAO method \cite{mullikenElectronicPopulationAnalysis1955}, the carrier-envelope offset frequencies and stability properties of the various extendons can be better-understood. First, the phase NLSE is rewritten as
\begin{equation}\label{eqn:Hrelation}
    i\frac{\partial F}{\partial T}=\widehat{H}F,
\end{equation}
where $\widehat{H}$ is the Hamiltonian of the phase NLSE, given by
\begin{equation}\label{eqn:hamiltonian}
    \widehat{H}=-\frac{1}{2}\beta\frac{\partial^2}{\partial z^2}-\gamma \left|F\right|^2\left(\arg{F}-\langle\arg{F}\rangle\right)-ir\left(\left|F\right|^2-P_0\right).
\end{equation}
At this point, the phase NLSE resembles a traditional time-evolution problem, albeit a nonlinear one.  It is therefore instructive to consider the eigenvalues, defined by the $F_N$ that satisfy
\begin{equation}\label{eqn:eigenvalue}
    \widehat{H}F_N=\lambda_N F_N.
\end{equation}
Functions that satisfy this equation will make up the stationary solutions or harmonic states of the system. The full solution can then be expressed as a sum of these stationary states multiplied by an appropriate complex exponential. Combining (\ref{eqn:Hrelation}) and (\ref{eqn:eigenvalue}), one gets
\begin{equation} \label{eqn:solsum}
     F\left(T,z\right)=\sum_{N}^\infty a_N F_N\left(z\right)e^{-i\lambda_N T},
\end{equation}
where $F_N$ is the Nth harmonic state and $a_N$ is a weighting factor. Note that for completeness one should also include the trivial $N=0$ state, the continuous-wave (CW) solution with $F=\sqrt{P_0}$.

Because harmonic states have the same chirp as fundamental states, they are also solutions to the NLSE, and can also be taken as eigenstates. However, in truth neither the fundamental extendon nor the harmonic extendon calculated above are the \textit{true} eigenstates, as they are not differentiable at the jump. Currently, we do not have a working form for the true eigenstates---they may not even be expressible in terms of elementary functions. Nevertheless, because the extendons previously proposed are very close to the true eigenstates in an integrated error sense, it is valuable to consider the expectation value of the Hamlitonian as the approximate eigenvalues.

To find an approximate expression for the $\lambda_N$, the diagonal elements of $\widetilde{H}$, the matrix corresponding to $\widehat{H}$, are calculated. We note that by following this procedure one obtains the same results compared to using the full matrices, but the interpretation is more straightforward. Finding these diagonal elements is a simple matter of calculating the inner product
\begin{equation} \label{eqn:Helt}
    \widetilde{H}_{NN}=\langle F_N \left|H\right| F_N\rangle=\int_{-L_c}^{L_c}F_N^\ast \left(HF_N\right)dz,
\end{equation}
and normalizing to the overlap
\begin{equation}\label{eqn:Selt}
    \widetilde{S}_{NN}=\langle F_N | F_N \rangle=\int_{-L_c}^{L_c} F_N^\ast F_N dz.
\end{equation}
Performing this calculation, one finds that 
\begin{equation} \label{eqn:eigval12}
    \lambda_{N}=\frac{\widetilde{H}_{NN}}{\widetilde{S}_{NN}}=\frac{2 L_c^2 r^2 \gamma^2 P_0^2}{3\beta\left(2r+\gamma\right)^2}\left(\frac{1}{N}+i\frac{3\beta}{\gamma A_0^2 L_c^2}\right).
\end{equation}
The imaginary component calculated above arises from the delta function that is generated by the discontinuiuty in the frequency at the boundary. One expects that the true eigenstates have imaginary parts that are exactly zero, but even if that is not the case they are numerically very small.
Thus, for an undisturbed system with no gain curvature, the eigenvalues have the form
\begin{equation} \label{eqn:eigvaln}
    \operatorname{Re}\lambda_N=\frac{\lambda}{N^2},
\end{equation} 
where $\lambda$ is used to denote the eigenvalue of the fundamental and $N$ is the index of the harmonic. This result holds true for all harmonic states of the form given in Eq. (\ref{eqn:hstate}). Referring to (\ref{eqn:solsum}), the evolution of each eigenstate has $e^{-i\lambda_N T}$ dependence, and so $\lambda_N$ has an additional interpretation as the carrier-envelope offset frequency of the extendon. Specifically, 
\begin{equation}\label{eqn:CEO}
    f_{N}^{\left(\textrm{ceo}\right)}=\frac{1}{2\pi}\frac{d}{dT}\arg{F_N}=-\frac{1}{2\pi}\lambda_N.
\end{equation}
Hence, the real part of these eigenvalues provides the CEO angular frequency its corresponding harmonic state.

Next, consider the imaginary component. This value is the constant for all of the harmonic states and manifests as a consequence of the discontinuity in the phase equation. This discontinuity ultimately arises from the boundary conditions present at the cavity's mirrors. The impact of this term is essentially to multiply each $F_N$ with a term which grows exponentially with $T$. From this, one would expect the harmonic states to be equally stable---but one can verify numerically that the lower-energy states are actually more robust. We will explore the implications of this analysis extensively in the subsequent sections, and show how defects can modify this behavior.

\subsection{Effect of gain curvature on stable extendons}
Although the results in the previous sections suffice for guiding intuition, they fail to account for the effects of gain curvature, a necessary consequence of the finite linewidth of a real laser. While it is known that gain curvature plays a crucial role in destabilizing combs at low dispersions, for stable combs its role is typically small. With gain curvature, the phase NLSE begins to more closely resemble the Complex-Ginzburg Landau equation, albeit one with an additional phase potential. In order to examine these effects, one can adjust many of the previous results for the phase-dominated NLSE by perturbatively modifying Eq. (\ref{eqn:NLSE}) and its related results.\\

\begin{figure}\centering
	\includegraphics[scale=0.5]{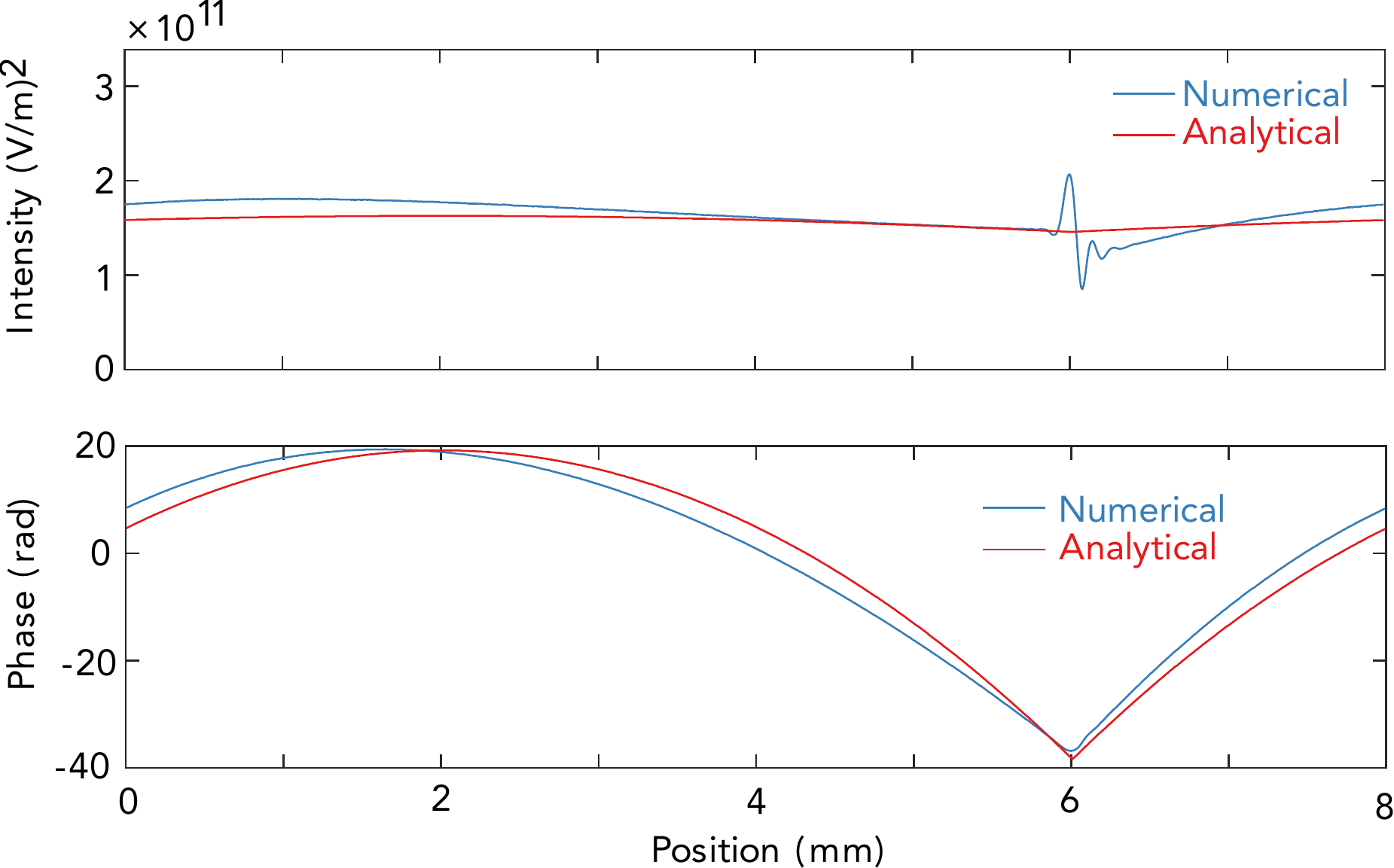}
	\caption{\label{fig:GainCurv} Gain curvature modifications (red) for the fundamental solution's intensity and phase compared with the numerical solution (blue).}
\end{figure}

\noindent To accomplish this goal, one essentially takes $\beta$ to be complex, i.e., replacing it with $\beta-\frac{i}{2}D_g$. Then, the phase NLSE (Equation (\ref{eqn:NLSE2})) is solved using a Gaussian ansatz
\begin{equation}\label{eqn:gain_fundamental}
    F_{t}(z,T)\equiv \sqrt{R_0 e^{-az^2}} e^{i\left(q z^2 -\lambda_t T\right)},
\end{equation}
where a is the decay factor of the Gaussian, $R_0$ is its amplitude, q is the chirp parameter, and $\lambda$ is the system's eigenvalue, related to the average phase. From here, one can perturbatively solve for the unknown variables $q$, $a$, $\lambda$, and $R_0$. After series expansions around $D_g=0$ and $\gamma=0$ and algebra, these can be approximated as:
\begin{align}\label{eqn:gain_qfactor}
    q&=\frac{\gamma A_0^2  }{2 \beta }+D_g \frac{A_0^4 \gamma ^3 L_c^2 }{24 \beta ^3 r} \\
    a&=D_g\frac{A_0^2 \gamma ^2}{4 \beta ^2 r} \\
    R_{0}&=A_0^2-D_g\frac{A_0^4 \gamma ^3 L_c^2}{24 \beta ^2 r^2} \\
    \lambda_t&= \frac{A_0^4 \gamma ^2 L_c^2}{6 \beta }-D_g\frac{A_0^2 \gamma}{4 \beta } 
\end{align}
The details are long and are contained in a supplemental notebook \cite{MathematicaFile}. The leftmost term in the above expressions represents the equivalent term unmodified by gain curvature. While only approximate, these terms are a reasonable approximation for moderate values of $D_g$, as demonstrated in Fig. \ref{fig:GainCurv}. 
The most important effect of gain curvature is to introduce a non-zero amplitude decay, which causes frequency modulation to induce amplitude modulation, decreasing intensity away from the center of the FM. It also modifies the chirp, peak power, and eigenvalue by a few percent.

The extension to harmonic extendons is straightforward. The eigenvalues are modified by a factor of $N^2$, leaving
\begin{equation}\label{eqn:gain_lambda_n}
    \begin{aligned}
    \operatorname{Re} \lambda_{tN} \approx {} & \frac{1}{N^2} \left(\frac{A_0^4 \gamma ^2 L_c^2}{6 \beta }-D_g\frac{A_0^2 \gamma}{4 \beta } \right),
    \end{aligned}
\end{equation}
while the chirp bandwidth $BW=q\frac{2L_c c}{\pi n}$ is reduced by N, leading to
\begin{equation}\label{eqn:chirpBW_Gain}
    BW_t = \left( \frac{\gamma A_0^2 }{2 \beta }+D_g \frac{A_0^4 \gamma ^3 L_c^2 }{24 \beta ^3 r}\right)\frac{2L_cc}{\pi n N}.
\end{equation}
For ease of reference, these results are summarized in Table \ref{tab:KeyResults}.

\renewcommand{\arraystretch}{2}
\arrayrulecolor{black}
\begin{center}
\begin{table}[!h] 
\centering
\begin{tabular}{||c|c|c||} 
 \hline
 \multicolumn{3}{|c|}{\bfseries Key extendon parameters with gain curvature} \\  
 \hline
 Intensity
    & $|F_t|^2$
    & $\left[A_0^2-D_g\frac{A_0^4 \gamma ^3 L_c^2}{24 \beta ^2 r^2} \right]\exp\left(-D_g\frac{A_0^2 \gamma ^2}{4 \beta ^2 r}z^2\right)$ \\ 
 \hline
 Phase 
    & $\arg F_t$ 
    & $\left(\frac{\gamma A_0^2  }{2 \beta }+D_g \frac{A_0^4 \gamma ^3 L_c^2 }{24 \beta ^3 r} \right)z^2$ \\ 
 \hline
 \makecell{Carrier-envelope\\offset frequency} 
    & $-\frac{\lambda_{tN}}{2\pi}$ 
    & $- \frac{1}{2\pi N^2} \left(\frac{A_0^4 \gamma ^2 L_c^2}{6 \beta }-D_g\frac{A_0^2 \gamma}{4 \beta } \right)$ \\
 \hline
 Harmonic states 
    & $F_{tN}$
    & $F_t\left(z\bmod \frac{2L_c}{N},\frac{T}{N^2}\right)$ \\
 \hline
 Chirp bandwidth 
    & $BW_t$
    & $\left( \frac{\gamma A_0^2 }{2 \beta }+D_g \frac{A_0^4 \gamma ^3 L_c^2 }{24 \beta ^3 r}\right)\frac{2L_cc}{\pi n N}$ \\ 
 \hline
\end{tabular}
\caption{Summary of key extendon parameters, including intensity, phase, CEO frequencies, harmonic states, and chirp bandwidth.}
\label{tab:KeyResults}
\end{table}
\end{center}

\subsection{Analysis of reflective defects} \label{sec:defect}

The phase NLSE provides a natural framework for analyzing perturbations to simple cavities. Here, we explore the impact of adding reflective defects, which can create asymmetrical cavities and allow one to engineer the phase profile. This type of defect engineering was recently demonstrate using ring laser FM combs \cite{kazakovDefectengineeredRingLaser2021}, where it was shown that defects can induce a laser to act in a harmonic state (if desired). It is expected that similar behavior can be induced in Fabry-Perot cavities---in particular, we will demonstrate how adding a minor reflection defect in the center of the cavity causes the system to prefer the second harmonic state over the first.

To construct the effect of a minor defect inside our cavity, one can form a scattering matrix, $\widetilde{S}$, relating the incident, reflected, and transmitted fields to each other \cite{pozarMicrowaveEngineering2004a}. We use this information to formulate an appropriate field term for an arbitrary collection of small reflection defects (see Appendix \ref{sec:refldefectterm} for the full derivation). The resulting master equation term is
\begin{equation} \label{eqn:defect_term}
    \left(\frac{n}{c}\frac{\partial{E_\pm}}{\partial{t}}+\frac{\partial E_\pm}{\partial z}\right)_\textrm{defect}=\sum_j \frac{1}{2} \ln \left( \frac{1+r_j}{1-r_j}\right)\delta\left(z-z_j\right) E_\mp,
\end{equation}
where $r_j$ is the field reflectivity of the defects and $z_j$ are their physical locations within the cavity. Converting to the normalized and extended cavity, one finds that 
\begin{equation}\label{eqn:defect_term_F}
    \left(\frac{n}{c}\frac{\partial{F}}{\partial{t}}+\frac{\partial F}{\partial z}\right)_\textrm{defect}=\sum_j \frac{1}{2}\ln\left(\frac{1+r_j}{1-r_j}\right)\left[\delta\left(z+z_j \right)+\delta\left(z-z_j \right)\right]\frac{K_-^{1/2}\left(z\right)}{K^{1/2}\left(z\right)}F_-\left(z\right),
\end{equation}
where the defect now appears twice to account for the fact that waves see it twice a round trip. Performing the mean-field averaging and employing the spatiotemporal substitution $ \partial t\rightarrow\ -n/c\ \partial z$, one find that
\begin{equation}\label{eqn:integrated_defect_term}
    \left(\frac{\partial F}{\partial T}\right)_{\textrm{defect}}=\frac{c}{2L_cn}\sum_j\frac{1}{2} \ln\left(\frac{1+r_j}{1-r_j}\right)\left[\frac{K_-^{1/2}\left(-z_j\right)}{K^{1/2}\left(-z_j\right)}F\left(z+2z_j \right)+\frac{K_-^{1/2}\left(z_j \right)}{K^{1/2}\left(z_j\right)}F\left(z-2z_j\right)\right],
\end{equation}
since the delta functions give rise to a shift.
Note that the defect derivation is only valid when $|r_j|\ll 1$ if one wishes to make use of the above results for steady-state intensity, as a sufficiently large reflector will begin to violate the mean-field assumption of small internal changes. To make it valid for large values, one would need to directly modify the effective gain upon which $K\left(z\right)$ is based. This is necessary to account for the additional reflections. The impact of small $r_j$ on the effective gain is negligible. Thus,  (\ref{eqn:integrated_defect_term}) provides a useful explanation of the dynamics added by a reflection defect.\\

\begin{figure}\centering
	\includegraphics[scale=0.7]{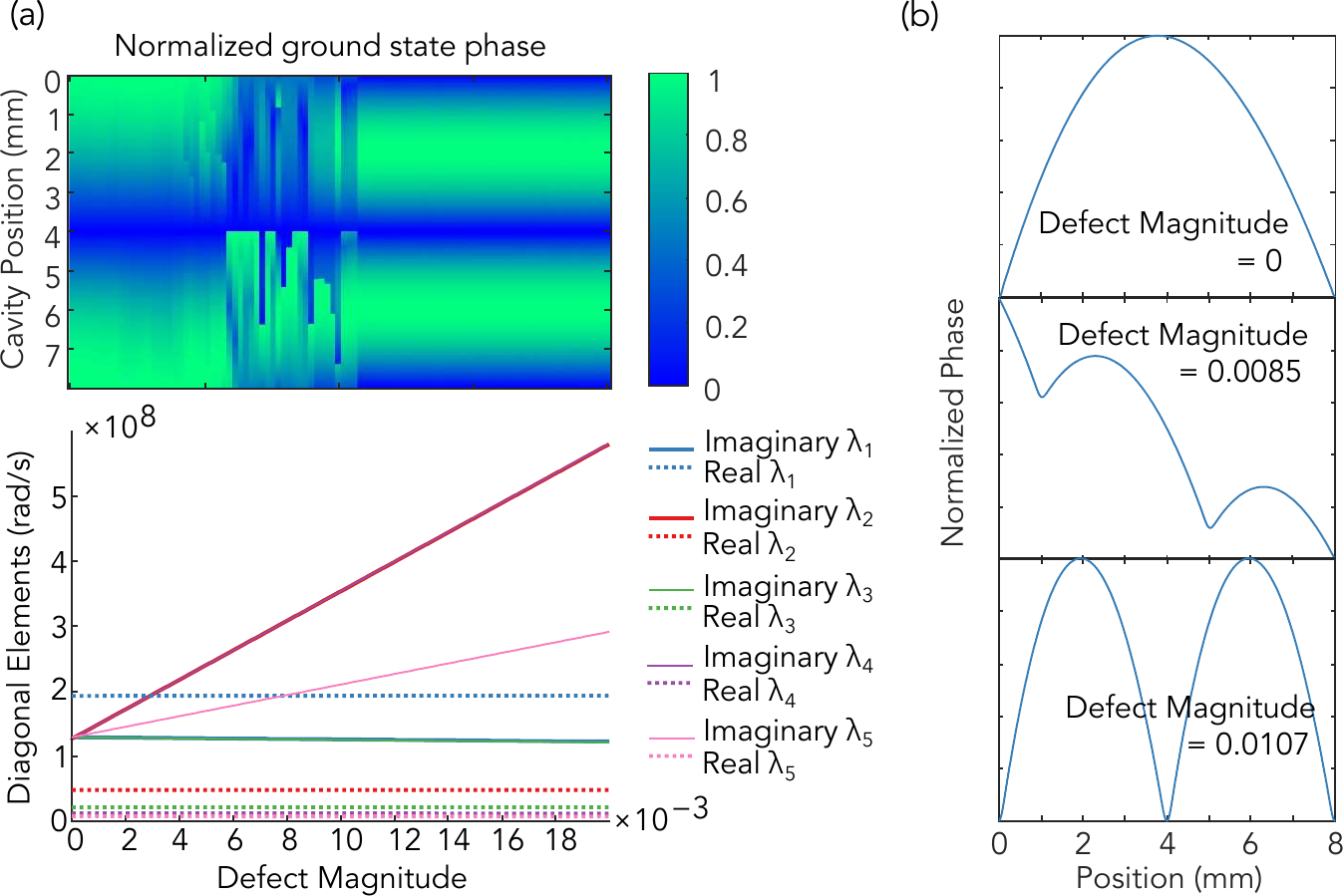}
	\caption{\label{fig:Defect}(a) Top: phase heatmap showing the transition between the fundamental harmonic and the second harmonic, as a function of defect size. These values were obtained after thousands of simulated round-trips; for moderate defect sizes, the system becomes unstable. Bottom: calculated diagonal elements corresponding to the first five harmonic states.
	(b) Phase plots at specific defect magnitudes ($r_1$ values)}
\end{figure}

Figure \ref{fig:Defect} shows the effects of a defect term on the ground state solution of a system with a defect located at $L_c/2$ and magnitude $r_1$. Applying Eq. (\ref{eqn:integrated_defect_term}), this leads to mean-field terms of the form $F\left(z-L_c\right)$ and $F\left(z+L_c\right)$, essentially coupling fields separated by a distance of $L_c$. As this matches the peak spacing for the second harmonic of the system (Fig. \ref{fig:Harmonic}) but not the other harmonics, this leads to the second harmonic becoming the new ground state solution as the defect magnitude increases. It is worth noting that in between the regions where the first harmonic and the second harmonic are dominant, quasi-stable solutions still exist despite the chaotic appearance of the heat plot in that region (Fig. \ref{fig:Defect}a). However, these solutions are not time invariant; they are breather states that fluctuate periodically with time. The middle plot of Figure \ref{fig:Defect}b shows the phase profile of one of these states at a fixed time.\\

To provide a quantitative explanation of this behavior, the eigenvalue approach from Section \ref{sec:eigenstates} can be employed. Adding the following defect operator to the Hamiltonian established in (\ref{eqn:hamiltonian}), one obtains
\begin{equation}\label{eqn:defect_Hop}
    H_\textrm{defect} = i \frac{c}{2L_cn}\sum_j \frac{1}{2} \ln\left(\frac{1+r_j}{1-r_j}\right)\left[\frac{K_-^{1/2}\left(-z_j \right)}{K^{1/2}\left(-z_j \right)}T^{2z_j}+\frac{K_-^{1/2}\left(z_j\right)}{K^{1/2}\left(z_j\right)}T^{-2z_j}\right],
\end{equation}
where $T^t=e^{t \partial/\partial z}$ is the shift operator $\left(T^tf\left(z\right)=f\left(z+t\right)\right)$. For the example detailed in Fig. \ref{fig:Defect}, (\ref{eqn:defect_Hop}) reduces to 
\begin{equation}\label{eqn:defect2_Hop}
    H_\textrm{defect} = i \frac{c}{2L_cn}\frac{1}{2} \ln\left(\frac{1+r_1}{1-r_1}\right)\left[\frac{K_-^{1/2}\left(-\frac{L_c}{2}\right)}{K^{1/2}\left(-\frac{L_c}{2} \right)}T^{L_c}+\frac{K_-^{1/2}\left(\frac{L_c}{2}\right)}{K^{1/2}\left(\frac{L_c}{2}\right)}T^{-L_c}\right].
\end{equation}
Denote the positive constants attached to $T^{L_c\ }$ and $T^{-L_c\ }$ as $C_+$ and $C_-$, respectively. The defect term’s contribution to $\widetilde{H}_{NN}$ is
\begin{align}\label{eqn:Helt_defect}
    \left(\widetilde{H}_{NN}\right)_{\textrm{defect}}&=\langle F_N \left|i\left(C_+ T^{L_c}+C_- T^{-L_c}\right)\right| F_N \rangle \nonumber \\
    &=iC_+ \langle F_N\left(z\right)\left|F_N\left(z+L_c\right)\rangle+iD_+ \langle F_N\left(z\right)\right|F_N\left(z-L_c\right)\rangle.
\end{align}
Note that for even harmonic states (denoted by $F_e$), it must be the case that $F_e\left(z\right)=F_e\left(z\pm L_c\right)$. Therefore, it must be the case that
\begin{equation}\label{eqn:Helt_defect_even}
    \left(\widetilde{H}_{ee}\right)_{\textrm{defect}}=i2L_c\left(C_+ + C_- \right)\left|F\right|^2,
\end{equation}
where $\left|F\right|^2$ is the amplitude squared of the harmonic state. By contrast, for odd diagonal elements, these matrix elements will be complex rather than purely imaginary. The amplitude of all the harmonic states is the same and the overlap is much worse for all other state configurations. Hence, it follows that the even index diagonal elements of $\widetilde{H}$ will receive the dominant contribution from the defects, as shown in Fig. \ref{fig:Defect}a. In other words,
\begin{equation}\label{eqn:H_ineq}
    \operatorname{Im}\left(\widetilde{H}_{ee}\right)_{\textrm{defect}}>\operatorname{Im}\left(\widetilde{H}_{oo}\right)_{\textrm{defect}},
\end{equation}
where $\widetilde{H}_{oo}$ represents the eigenvalue change of odd diagonal elements. For sufficiently small reflectivity, $\frac{1}{2} \ln\left(\frac{1+r_1}{1-r_1}\right) \approx r_1$, so an approximate shift of the eigenvalue for an even state will be given by
\begin{equation}\label{eqn:defect_lambda}
    \lambda_{\textrm{defect}}^{\left(e\right)}=i \frac{c}{n} r_1 \left|F\right|^2\left(\frac{K_-^{1/2}\left(-\frac{L_c}{2}\right)}{K^{1/2}\left(-\frac{L_c}{2}\right)}+\frac{K_-^{1/2}\left(\frac{L_c}{2}\right)}{K^{1/2}\left(\frac{L_c}{2}\right)}\right).
\end{equation}
This perturbation is purely imaginary, and so it will generate an evolution of $e^{-i \lambda_{\textrm{defect}}^{\left(e\right)} T}$. Because positive imaginary part corresponds to exponential growth, 
the presence of defects now favors even states. Once a defect is turned on, even states grow faster than odd states, causing them to be dominant. In steady-state a perturbation will always develop at the boundary to cancel this growth out; this perturbation is energetically easiest for the lowest-order state, and so the system favors the $N=2$ state.


The eigenvalue approach utilizing the harmonic basis set is quite informative, however, it does not fully explain why the system always prefers the first harmonic as its ground state solution in the absence of defects or why the ground state solution prefers to the second harmonic over other even states. The explanation is that the current formalism fails to adequately account for the boundary spike, which lifts the degeneracy of the eigenstates. Numerical investigation has revealed that a system initialized in a higher-order state always decays to the lowest possible harmonic (for an unmodified system, this would be the fundamental solution). Adding reflection defects changes this base system to prefer the lowest even state ($N=2$). Thus, to make more detailed predictions analytically, one would also need to account for boundary spike in the eigenstate. Of course, stability information can still be obtained numerically.

\section{Conclusion}
We have performed a comprehensive analytical treatment of FM comb behavior. We demonstrated how one can generalize active cavity mean-field theory to handle various laser systems, including systems with defects or other perturbations. We described how one can derive a phase-driven NLSE for a cavity with different facet reflectivities and derived equations for several key parameters in terms of known physical constants, and showed how it can be modified to account for nonidealities such as gain curvature. Finally, we showed how a formulation of the phase-dominated NLSE as an eigenvalue problem, where harmonic states represent the higher-order eigenstates, allows one to analyze the stability these systems in the presence of defects.

\medskip
\section{Disclosures}
\medskip
The authors declare no conflicts of interest.

\medskip
\section{Acknowledgments}
\medskip
The authors acknowledge support by the Department of Defense (DoD) through the National Defense Science and Engineering Graduate (NDSEG) Fellowship Program, the Air Force Office of Scientific Research Young Investigator Program (award number FA9550-20-1-0192), the National Science Foundation CAREER program (award number ECCS-2046772), and the Office of Naval Research Young Investigator Program (award number N00014-21-1-2735).

\clearpage
\appendix
\section{Simulation parameters and values}
For completeness, all simulation parameters used inside this work are included in Table \ref{tab:SimValues}.
\renewcommand{\arraystretch}{1.5}
\arrayrulecolor{black}
\begin{center}
\begin{table}[!htbp] 
\centering
\vspace*{-5mm}
\begin{tabular}{|c c c|}
\hline
{\bfseries Name} & {\bfseries Symbol} & {\bfseries Value}\\ 
\hline
Population lifetime & $T_1$ & $0.4 \mbox{ } \textrm{ps}$\\[0ex]
Coherence lifetime & $T_2$ & $50 \mbox{ } \textrm{fs}$\\
Refractive index & $n$ & $3.3$ \\ 
Waveguide losses & $\alpha_w$ & $4 \mbox{ } \textrm{cm}^{-1}$\\
Dipole moment & $z_0$ & $2.3 \mbox{ } \textrm{nm}$ \\
Left mirror reflectivity & $R_1$ & variable \\
Right mirror reflectivity & $R_2$ & variable\\
Cavity length & $L_c$ & $4 \mbox{ } \textrm{mm}$\\
Module length & $L_\textrm{mod}$ & $58 \mbox{ } \textrm{nm}$\\
Wavelength & $\lambda_0$ & $8 \mbox{ } \mu \textrm{m}$\\
Current density & $J$ & $1100 \mbox{ } \frac{\textrm{A}}{\textrm{cm}^2}$\\
Dispersion & $k''$ & $-2000 \mbox{ } \frac{\textrm{fs}^2}{\textrm{mm}}$\\
Kerr nonlinearity & $\gamma_K$ & $0$\\
Reflection defect & $r_1$ & variable\\
Reflection defect position & $z_1$ & $2$ mm\\
\hline
\end{tabular}

\caption{Parameters of a two-level system used during numerical analysis.}
\label{tab:SimValues}
\end{table}
\end{center}

\section{Derivation of reflection defect term}\label{sec:refldefectterm}
In this section, we show how a small reflective defect can be formulated in a spatially-extended way that is suitable for mean-field theory. For master equations the most straightforward approach is to use explicit boundary conditions, but for mean-field theory these are difficult to integrate. The essential idea is to convert the boundary conditions into a gain \textit{matrix} that can act on both master equations at once. Consider the situation illustrated in Fig. \ref{fig:DefectSetup}, from this setup, we can form a scattering matrix, $\widetilde{S}$, relating the fields (the four shown in the figure) to each other as follows \cite{hausWavesFieldsOptoelectronics1983a}:
\begin{figure}\centering
	\includegraphics[scale=0.6]{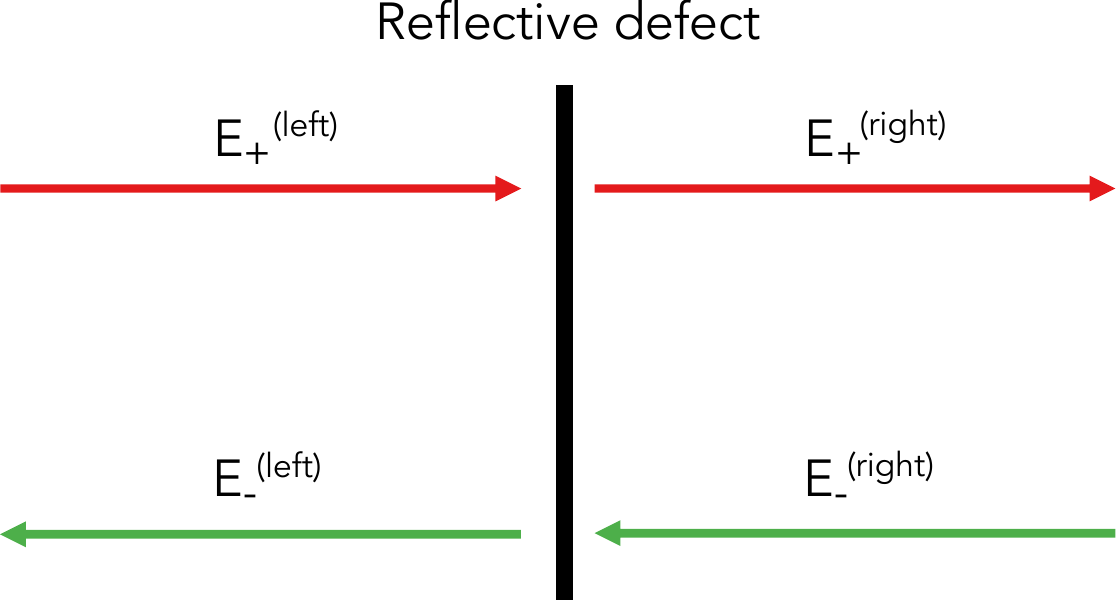}
	\caption{\label{fig:DefectSetup} Forward and backward traveling waves encountering a defect}
\end{figure}
\begin{equation}\label{eqn:field_M_equation}
\begin{pmatrix}
E_-^{\left(\textrm{left}\right)}\\
E_+^{\left(\textrm{right}\right)}
\end{pmatrix}
=\widetilde{S}
\begin{pmatrix}
E_-^{\left(\textrm{right}\right)}\\
E_+^{\left(\textrm{left}\right)}
\end{pmatrix}
\end{equation}
where
\begin{equation}\label{eqn:scatteringM}
\widetilde{S} = 
\begin{pmatrix}
r & it\\ 
it & r
\end{pmatrix}.
\end{equation}
Here the defect reflection is given as $r$ and the transmission is given as $t$. Without loss of generality, one can shift the reference plane of the scattering matrix by multiplying left and right \cite{pozarMicrowaveEngineering2004a} by
\begin{equation}\label{eqn:PhaseAdjustMat}
\widetilde{P} = 
\begin{pmatrix}
-i & 0\\ 
0 & 1
\end{pmatrix},
\end{equation}
resulting in
\begin{equation}\label{eqn:Mprime}
\widetilde{S}'=\widetilde{P}\widetilde{S}\widetilde{P}= 
\begin{pmatrix}
-r & t\\ 
t & r
\end{pmatrix}.
\end{equation}
Next, this matrix is converted from a scattering matrix into a transmission matrix, in order to represent the change that occurs going from the left fields to the right fields. This is accomplished by solving for $E\pm^{\left(\textrm{left}\right)}$ in terms of $E\pm^{\left(\textrm{right}\right)}$, resulting in an expression that relates the fields on one side of the defect to the fields on the other side:
\begin{equation}\label{eqn:ScatterTransform}
\begin{pmatrix}
E_+^{\left(\textrm{right}\right)} \\ 
E_-^{\left(\textrm{right}\right)}
\end{pmatrix}
=
\begin{pmatrix}
\frac{1}{t} & \frac{r}{t}\\ 
\frac{r}{t} & \frac{1}{t}
\end{pmatrix}
\begin{pmatrix}
E_+^{\left(\textrm{left}\right)} \\ 
E_-^{\left(\textrm{left}\right)}
\end{pmatrix}
\equiv\widetilde{T}\begin{pmatrix}
E_+^{\left(\textrm{left}\right)} \\ 
E_-^{\left(\textrm{left}\right)}
\end{pmatrix}
\end{equation}
Defining the gain matrix $\widetilde{G}(z)$ as the matrix that satisfies
\begin{equation}\label{eqn:G_equation}
\frac{\partial}{\partial z}
\begin{pmatrix}
E_+\\
E_-
\end{pmatrix}
=\frac{1}{2}\widetilde{G}(z)
\begin{pmatrix}
E_+\\
E_-
\end{pmatrix},
\end{equation}
one finds that the formal solution is
\begin{equation}\label{eqn:G_IntEq}
\begin{pmatrix}
E_+(z_j+\epsilon)\\
E_-(z_j+\epsilon)
\end{pmatrix}
=\exp{\left(\frac{1}{2}\int_{z_j-\epsilon}^{z_j+\epsilon}\widetilde{G}(z')dz'\right)}
\begin{pmatrix}
E_+(z_j-\epsilon)\\
E_-(z_j-\epsilon)
\end{pmatrix}.
\end{equation}
Equating the exponential on the right-hand side of Eq. (\ref{eqn:G_IntEq}) with the transmission matrix $\widetilde{T}$, one can use the matrix logarithm to find
\begin{equation}\label{eqn:Equate_G_T}
\log \left(\widetilde{T}\right) = \frac{1}{2}\int_{z_j-z}^{z_j-z}\widetilde{G}(z')dz' 
=\begin{pmatrix}
0 &  \frac{1}{2}\log \left(\frac{1+r}{1-r} \right)\\
\frac{1}{2}\log \left(\frac{1+r}{1-r} \right) & 0
\end{pmatrix}.
\end{equation}
From this expression, it is clear that to satisfy the relationship between $\widetilde{G}$ and $\log \left(\widetilde{T}\right)$, one must take the defect's contribution to the gain as
\begin{equation}\label{eqn:G_logT}
\widetilde{G}_\mathrm{defect} =
\begin{pmatrix}
0 &  \log \left(\frac{1+r}{1-r} \right) \delta \left(z-z_j\right)\\
\log \left(\frac{1+r}{1-r} \right) \delta \left(z-z_j\right) & 0
\end{pmatrix}.
\end{equation}
This causes counterpropagating waves to be coupled by a factor of $\log \left(\frac{1+r}{1-r} \right) \delta \left(z-z_j\right)$, leading to the master equation
\begin{equation} \label{eqn:defect_term2}
    \left(\frac{n}{c}\frac{\partial{E_\pm}}{\partial{t}}+\frac{\partial E_\pm}{\partial z}\right)_\textrm{defect}=\sum_j \frac{1}{2} \ln \left( \frac{1+r_j}{1-r_j}\right)\delta\left(z-z_j\right) E_\mp.
\end{equation}

Finally, note that one must take some caution when constructing mean-field theory. Within the extended cavity formalism, defects actually appear twice---once for each pass through the cavity.

\section{Effect of additional third-order nonlinearities: nonlinear group velocity shifts}\label{sec:nonlingv}

Previous analysis \cite{burghoffSupplementaryDocumentFrequencymodulated2020} has discussed additional self-steepening terms as being a source of group delay, but this has not been studied as they are not key to understanding FM comb formation. However, they are worthy of investigation because they are a potential explanation for a commonly encountered discrepancy between the linear free-spectral range (FSR) of a cavity and the optimal frequency for injection locking \cite{hillbrandCoherentInjectionLocking2019a}. To that end, we will derive an expression estimating this group delay in an idealized situation without gain curvature. The additional nonlinear and linear terms that can induce group velocity shifts are
\begin{align}\label{eqn:gd_terms}
    \left(\frac{\partial F}{\partial T}\right)_{\textrm{gd}}={} & \frac{g_0}{2} T_2 \left(\frac{c}{n}\right)^2 \frac{\partial F}{\partial z}-\frac{g_0}{2P_s} \left[\left(T_1+\frac{5}{2} T_2 \right)\left( \widetilde{K}\left[\left|F\right|^2\right]+
    \langle K \rangle \left|F\right|^2 \right) \frac{\partial F}{\partial z}\right.\nonumber\\
    &\left.\hspace{3cm} +\left(T_1+\frac{3}{2} T_2 \right) \langle K \rangle F^2 \frac{\partial F^\ast}{\partial z}\right],
\end{align}
where the first term is the linear shift due to the system's finite linewidth, and the latter three are the additional third-order nonlinearities. To find the group delay caused by the terms in (\ref{eqn:gd_terms}), we need to get the prefactor attached to $ \partial F/\partial z $ from each of these terms and average them over the whole round trip. To perform these averages, we will assume $F\left(z\right)$ is in the fundamental state. Then, all the averages from the first line of Eq. (\ref{eqn:gd_terms}) are trivial to perform. Averaging the second line is more subtle though, so we will describe the steps in detail.
\begin{figure}\centering
	\includegraphics[width=1\linewidth]{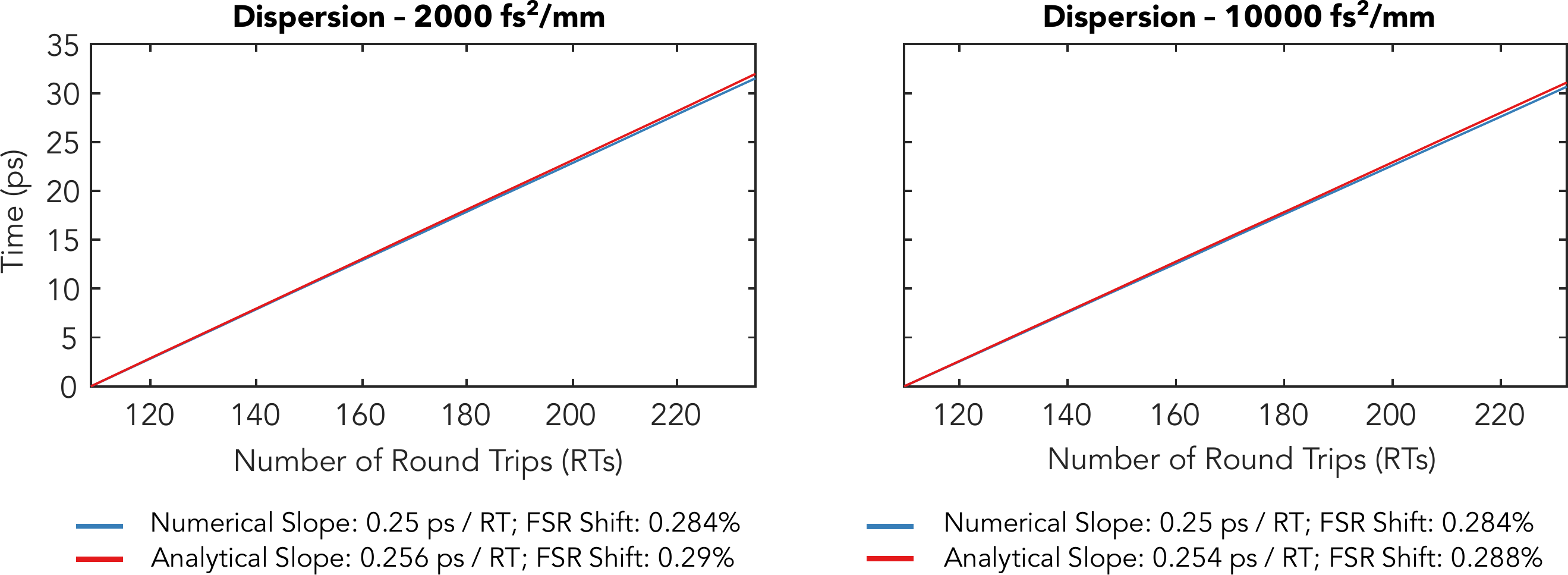}
	\caption{\label{fig:GD}Analytically calculated group delay from self-steepening terms and refractive index shift (red) compared to numerically simulated group delay using the same terms. The numerical version is not a function of dispersion, but the analytical version is slightly impacted. The shift in the system’s free spectral range (FSR) is also calculated.}
\end{figure}
First, note that
\begin{equation}\label{eqn:product_rule_F}
    F \frac{\partial F^\ast}{\partial z}=\frac{\partial\left|F\right|^2}{\partial z}-F^\ast \frac{\partial F}{\partial z}.
\end{equation}
Now using Eq. (\ref{eqn:product_rule_F}) to rewrite the second line of Eq. (\ref{eqn:gd_terms}) we obtain
\begin{equation}\label{eqn:ampl_dep_gd}
    \left(T_1+\frac{3}{2} T_2\right) \langle K \rangle F\left(\frac{\partial\left|F\right|^2}{\partial z}-F^\ast \frac{\partial F}{\partial z}\right)=\left(T_1+\frac{3}{2} T_2\right) \langle K \rangle \left(F \frac{\partial\left|F\right|^2}{\partial z}-\left|F\right|^2 \frac{\partial F}{\partial z}\right).
\end{equation}

In Eq. (\ref{eqn:ampl_dep_gd}) we have a term with an explicit dependence on $ \partial F/\partial z $ which can be treated the same as the other terms in Eq. (\ref{eqn:gd_terms}). The $ \left(\partial\left|F\right|^2\right)/\partial z $ term on the other hand is still problematic. It can be thought of as an intensity (or amplitude) induced group delay source unconnected to the phase. For systems with negligible gain curvature the amplitude is practically constant (save the boundary spike) and hence, the adding the averaged results provides a reasonably accurate description of the group delay associated with these terms (see Fig. \ref{fig:GD}). However, when stronger gain curvature is present, the amplitude is curved in a quadratic fashion and this term can no longer be ignored. Due to its atypical form, it is difficult to analyze currently.\\

In the case of negligible gain curvature as discussed above, the group delay per round trip associated with the terms specified in Eq. (\ref{eqn:gd_terms}), denoted by $T_d$, can be approximated as
\begin{equation}\label{eqn:gd_perTr}
    T_d \approx \frac{g_0}{2P_{\textrm{sat}}}\left[T_2 P_{\textrm{sat}}-\left(T_1+\frac{7}{2} T_2\right)\left|F\right|^2 \frac{\langle P \rangle}{P_0}\right]\left(\frac{c}{n}\right).
\end{equation}
Eq. (\ref{eqn:gd_perTr}) gives the group delay in seconds. We also calculated the percent FSR shift caused by this change in group delay, defined as
\begin{equation}\label{eqn:gd_FSR}
    \Delta \mbox{FSR}=\frac{1}{T_r}-\frac{1}{T_r+T_d}
\end{equation}
The agreement between the theoretical and numerical results, shown in Figure \ref{fig:GD}, is quite good. While the numerical value of the fractional FSR shift is somewhat smaller than what has been reported (0.3\% here, versus the 2.7\% measured in Ref. \cite{hillbrandCoherentInjectionLocking2019a}), it is conceivable that additional mechanisms can contribute to these shifts.


\bigskip

\bibliography{OE_Paper_LeviHumbard}

\end{document}